%% file: main.tex
\documentclass[conference]{IEEEtran}
\usepackage{cite}
\usepackage{amsmath,amssymb,amsfonts}
\usepackage{graphicx}
\usepackage{textcomp}
\usepackage{xcolor}

\author{%
  \IEEEauthorblockN{%
    Nathan Gober\IEEEauthorrefmark{1}, Gino Chacon\IEEEauthorrefmark{1}, Lei Wang\IEEEauthorrefmark{1}, Paul V. Gratz\IEEEauthorrefmark{1},\\%
  Daniel A. Jim\'{e}nez\IEEEauthorrefmark{1}, Elvira Teran\IEEEauthorrefmark{2}, Seth Pugsley\IEEEauthorrefmark{3}, Jinchun Kim\IEEEauthorrefmark{1}%
  }
  \vspace{6pt}

  \IEEEauthorblockA{\IEEEauthorrefmark{1} Texas A\&M University\\%
  \{ngober, ginochacon, wilsonwang2019\}@tamu.edu, pgratz@gratz1.com, djimenez@acm.org, cienlux@gmail.com}
  \IEEEauthorblockA{\IEEEauthorrefmark{2} Texas A\&M International University\\elviraterangarcia@gmail.com}
  \IEEEauthorblockA{\IEEEauthorrefmark{3} Intel\\sethpugsley@gmail.com}
}

\begin{document}

\title{The \textbf{Champ}ionship \textbf{Sim}ulator: Architectural Simulation for Education and Competition}

\maketitle
\thispagestyle{plain}
\pagestyle{plain}

\input{abstract}

\input{intro}

\input{champsim}

\input{modules}
\input{application}

\input{related_works}

\input{future_work}

\input{conclusion}

\bibliographystyle{IEEEtran}
\IEEEtriggeratref{32}
\bibliography{ref}

\end{document}

%% file: abstract.tex
\begin{abstract}

  Recent years have seen a dramatic increase in the microarchitectural
  complexity of processors.  This increase in complexity presents a
  twofold challenge for the field of computer architecture.  First, no
  individual architect can fully comprehend the complexity of the
  entire microarchitecture of the core.  This leads to increasingly
  specialized architects, who treat parts of the core outside their
  particular expertise as black boxes.  Second, with increasing
  complexity, the field becomes decreasingly accessible to new
  students of the field.  When learning core microarchitecture, new
  students must first learn the big picture of how the system works in
  order to understand how the pieces all fit together.  The tools used
  to study microarchitecture experience a similar struggle.  As with
  the microarchitectures they simulate, an increase in complexity
  reduces accessibility to new users.

  In this work, we present ChampSim.  ChampSim uses a
  modular design and configurable structure to achieve a low barrier
  to entry into the field of microarchitecural simulation.  ChampSim has
  shown itself to be useful in multiple areas of research,
  competition, and education.  In this way, we seek to promote access
  and inclusion despite the increasing complexity of the field of
  computer architecture.

\end{abstract}

%% file: intro.tex
\section{Introduction}

Computer architecture has developed over the decades into a mature
field with strong specialization.  Processors have grown in
complexity, gaining more and more features as transistor density
continues to increase, and as each new iterative innovation builds
upon previous solutions.  An increasingly lengthy historical
perspective is required to fully appreciate the state of the leading
edge of processors.  Also, the trend of software tools used to assist
microarchitecture development has been towards large, complex, and
monolithic simulation environments.

Simultaneously, computer architecture education has become a core
component of undergraduate computer engineering and computer science
curricula.  A basic understanding of computer architecture topics such
as cache replacement and branch prediction is no longer restricted to
graduate level courses, but can easily find a place in an
undergraduate research project or classroom assignment.  While
precise tools used by experts provide a high level of correctness,
students need tools appropriate for their level of knowledge,
experience, and ability.  A large gap exists between the need for
arcane tools used by experts and the need for accessible
tools that can be used by novices.  In this gap should stand a tool that
strikes a balance between correctness and accessibility.

Computer hardware development is impeded by the long production
pipelines required to realize a design.  If a hardware design must
traverse its entire life cycle in order to be evaluated, it
may take many years to produce a well-tested design.  It is
furthermore infeasible to tape out many iterations of a design in
order to evaluate their comparative performance.  In order to reduce
costs and mitigate long production pipeline times, hardware architects
turn to higher-level software simulation of their designs.  Simulation
bridges the gap by enabling rapid comparisons between the designs,
giving an approximation of the system's performance.  Software
simulation of designs is, by its nature, far slower than the hardware
designs being evaluated, but allows for a much shorter
design-to-evaluation time.

The advancement of the field of computer architecture depends on
accurate and rapid simulation.  Simulator authors make tradeoffs
between evaluation time and simulation accuracy.  The landscape of
simulators today is wide, but few see broad usage.  Some simulators
are less concerned with evaluation time, and are more concerned with
simulation accuracy.  Others place greater emphasis on model
performance over a more detail-oriented approach.  Such simulators
find their balance in a short evaluation time.

In this work, we present ChampSim, a simulator
designed to promote innovative research, inclusive education, and
healthy competition in the field of computer architecture.  The ChampSim
simulator is derived from the simulation tools used in the Second Data
Prefetching Competition, which was held in conjunction with ISCA 2015.
ChampSim simulates a heterogeneous multicore system with an arbitrary
memory hierarchy, where each out-of-order core can be configured
arbitrarily.  It is intended to promote access in the rapidly-growing
field of computer architecture.  To accomplish this, we keep three key
principles in mind: Low startup time, broad applicability, and design
configurability.  These design principles have led to a simulator that
favors usability and rapid iteration.

This paper will discuss in detail the purpose, design, and
effectiveness of the ChampSim simulator, beginning with a discussion of
the guiding principles of the design in Section~\ref{sec:design},
details on the key features of ChampSim's modular architecture in
Section~\ref{sec:modules}, a history of ChampSim's application in the
field and a vision of its continuing place in
Section~\ref{sec:application}, and a list of future development plans
in Section~\ref{sec:future}.

%% file: champsim.tex
\section{The ChampSim Architectural Simulator} \label{sec:design} The
ChampSim simulator has its roots in the simulation environment used for
the Second Data Prefetching Competition~\cite{dpc2}.  In a competition
environment, it is valuable to have an easy-to-use environment to
encourage wide participation and a variety of novel submissions.
These accessibility principles have continued through the development
of ChampSim and have emerged into three guiding design principles: low
startup time, broad applicability, and high configurablilty.

\subsection{Low startup time}
While failure is an important part of learning, wrangling with a
highly complex simulator before even beginning the implementation of a
new idea can frustrate the learning process of any student or
beginning researcher.  With this insight in mind, we seek that a new user should be
able to download and compile ChampSim in a few minutes, create their
first design in a few hours, and perform new and meaningful computer
architecture research within a few weeks.  In each of the applications
for which ChampSim is intended, it is valuable that a user is able to
begin using the simulator quickly.  Furthermore, the runtime of the
simulation should be short enough to provide quick feedback for a
novice user.

Many general-purpose processors have a lot in common.  Designs are
pipelined, usually with a decoupled, in-order front end and an
out-of-order back end.  Many researchers are not seeking to modify
these basic design aspects, but have a particular element of the
design in mind that they intend to study or improve.  ChampSim presents a
selection of areas that commonly see research activity as configurable
modules: branch predictors, cache replacement policies, branch target
buffers, and both instruction and data prefetchers.  These modules
provide an intuitive interface into a larger system, allowing
designers to test new designs quickly and effectively, while affording
them the opportunity to not have to worry about the parts of the
system they are not studying.

Reference implementations of legacy modules, such as the GShare branch
predictor~\cite{mcfarling93gshare} or the next-line prefetcher, are
included with the simulator.  These reference implementations can be
used as starting points for new designs or as placeholders if the user
is not interested in modifying those particular modules.

\subsection{Broad Applicability}
A researcher seeking to perform hardware research should not be
expected to be broadly familiar with a variety of programming
languages or to track their changes over time.  Therefore, for the
sake of inclusion, we seek that a user should only need an entry-level understanding of
C++, the language in which ChampSim is written, to perform research using ChampSim.  The interface to an simulator
should be simple and present the user with a few meaningful choices
that map well onto their experience.  Each module is fundamentally
only the implementation of a few functions.

ChampSim is trace-driven, meaning that simulation is performed in two
stages.  First, the workload to be simulated is instrumented and run
offline.  The tracing instrumentation produces a digest of the
program's activity, called a trace. The tracing step can be performed
offline from the simulation step and the trace can be stored in a
repository and made available to users.  To test a simulated design,
the user selects a trace file as input.  The trace is streamed into
the program as a stand-in for actual program execution.  This strategy
sacrifices a modest amount of accuracy, particularly in how the
operating system interacts with the program, in favor of ease in
reproducing results and of speed of the model.  It is simpler for ChampSim
to read a decoded trace file than to execute an external program, and
given established, compiled repositories of traces, it is a helpful
abstraction to the user to remove another step of environment setup.

Users are able to generate their own program traces with the included
``\textit{tracer}'' tracing tool, included in the ChampSim package.  The
included tracer is built upon Intel PIN~\cite{luk2005pldi}, a
well-documented tool for instrumenting programs at runtime, though
other tool sets, such as DynamoRIO~\cite{Bruening00designand}, can be
used.  Alternately, instruction traces can be dumped from
execution-driven simulators such as
gem5 or
QFlex~\cite{binkert2011gem5,power2020gem5,villalonga2019flexible}.  The tracer inspects every
instruction the program runs and encapsulates each instruction into a
decoded format that includes the instruction pointer, branching
behavior, and which registers and memory locations form the input and
output operands for the instruction.  The concatenation of many
instructions forms the entire trace.  This trace format permits ChampSim
to run with low memory requirements, since the trace does not need to
be held in memory but can be streamed off the disk after inline
decompression.

\subsection{Design configurability}
ChampSim is capable of modeling a large variety of commodity
processors.  
A configuration file specifies many aspects of the modeled CPU core,
including frequency, cache configuration, re-order buffer size, load
and store queue sizes, widths for instruction fetch, decode, execution
and retire, and a variety of latencies for different components.  In
addition, ChampSim includes a DRAM system that models bank and bus
contention.  The trace format includes only virtual memory addresses,
so ChampSim simulates a page table and TLB hierarchy with arbitrary
mappings of virtual to physical pages.

Each cache must be configured with a prefetcher (to simulate no
prefetcher, there is an included ``do-nothing'' prefetcher) and a
replacement policy.  The cache interfaces with a read queue, a write
queue, and a prefetch queue.  Prefetches originating from the cache
level are placed in its prefetch queue and serviced with a lower
priority than the read queue.  When a prefetch misses the cache, it is
by default sent to the lower level's prefetch queue, though this is
configurable.  All TLB structures function identically to caches.

Each core model is configured with an instruction prefetcher (which
may be a do-nothing prefetcher), a branch predictor, and a branch
target buffer.  Each core must be configured with an instruction TLB
and cache, plus a data TLB and cache.

The cache architecture is flexible.  The sink of all TLB requests must
be configured to be a core's hardware page table walker, whose lower
level is, in turn, the same core's L1 data cache, and the ultimate
sink of all instruction or data memory requests must be configured to
be the physical memory.  Beyond that, the cache hierarchy can be
arbitrary.  Caches can be shared between cores and can be non-uniform
levels of depth.

%% file: modules.tex
\begin{figure}[t]
    \centering
    \includegraphics[width=\columnwidth]{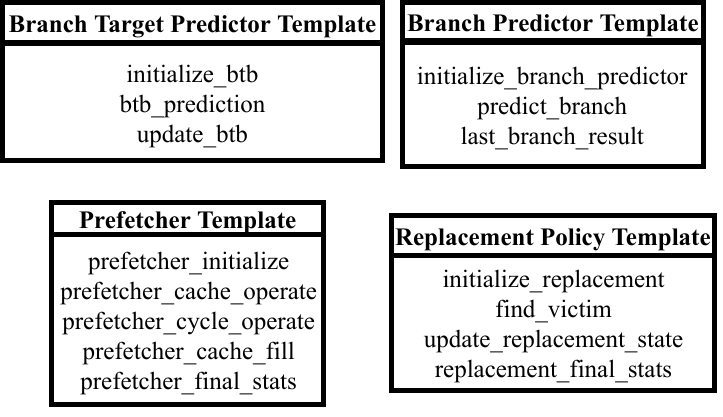}
    \caption{ChampSim defines four types of modules. Each module
      requires a specific set of functions, or hooks, to be defined to
      allow the module to interact with the underlying model.}
    \label{fig:templates}
\end{figure}

\begin{figure}[t]
    \centering
    \includegraphics[width=\columnwidth]{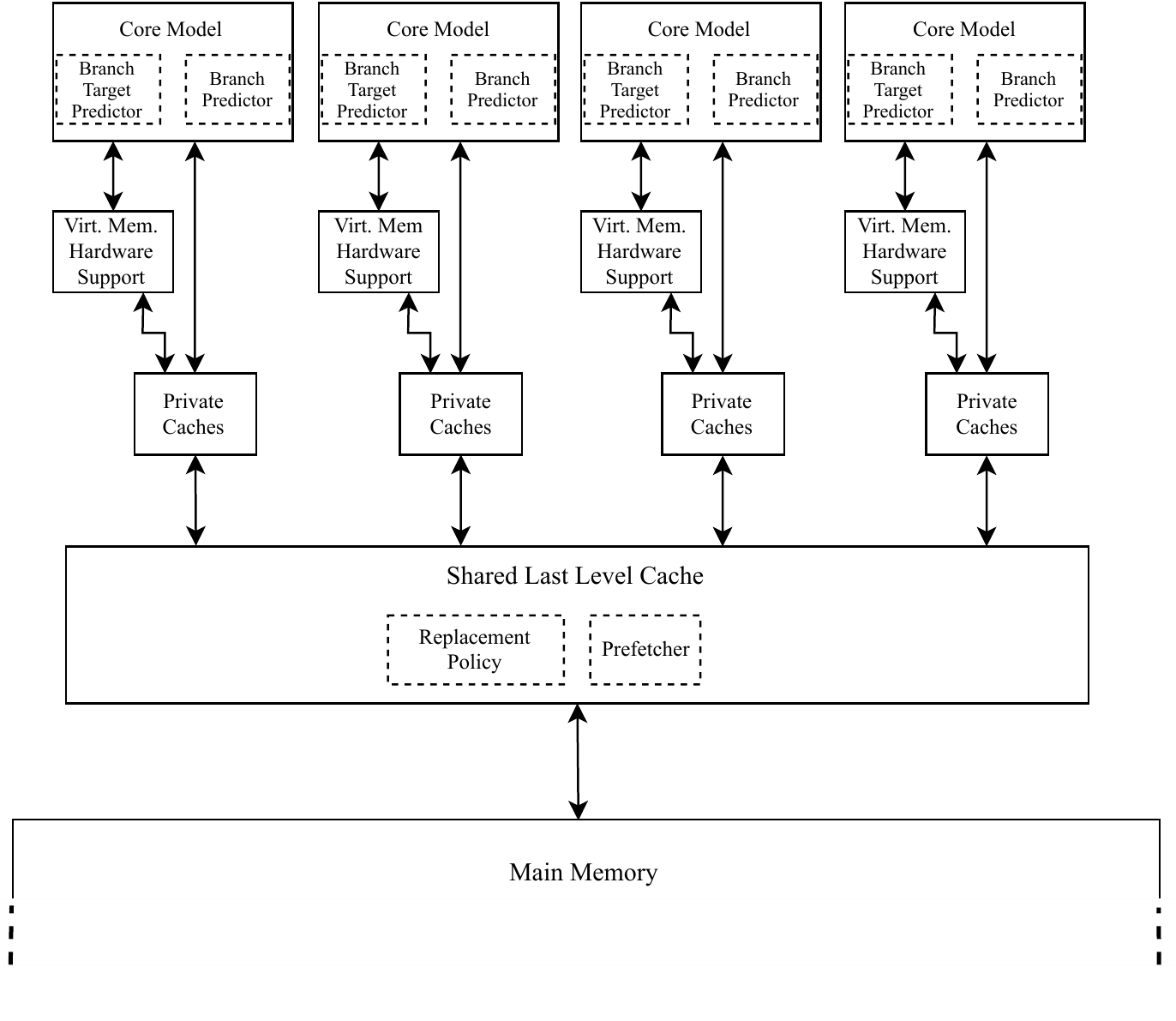}
    \hrule
        \caption{ChampSim is capable of simulating multicore
          systems, an example of which is given here. Each core has its own branch target predictor and
          branch prediction module that may be configured uniquely
          from one another. Each core has a private cache hierarchy
          and virtual memory hardware support. Every core's memory
          hierarchy shares a last level cache, which is subsequently
          connected to the main memory model.}
    \label{fig:modularity}
\end{figure}

\begin{figure}[t]
    \centering
    \includegraphics[width=.6\columnwidth]{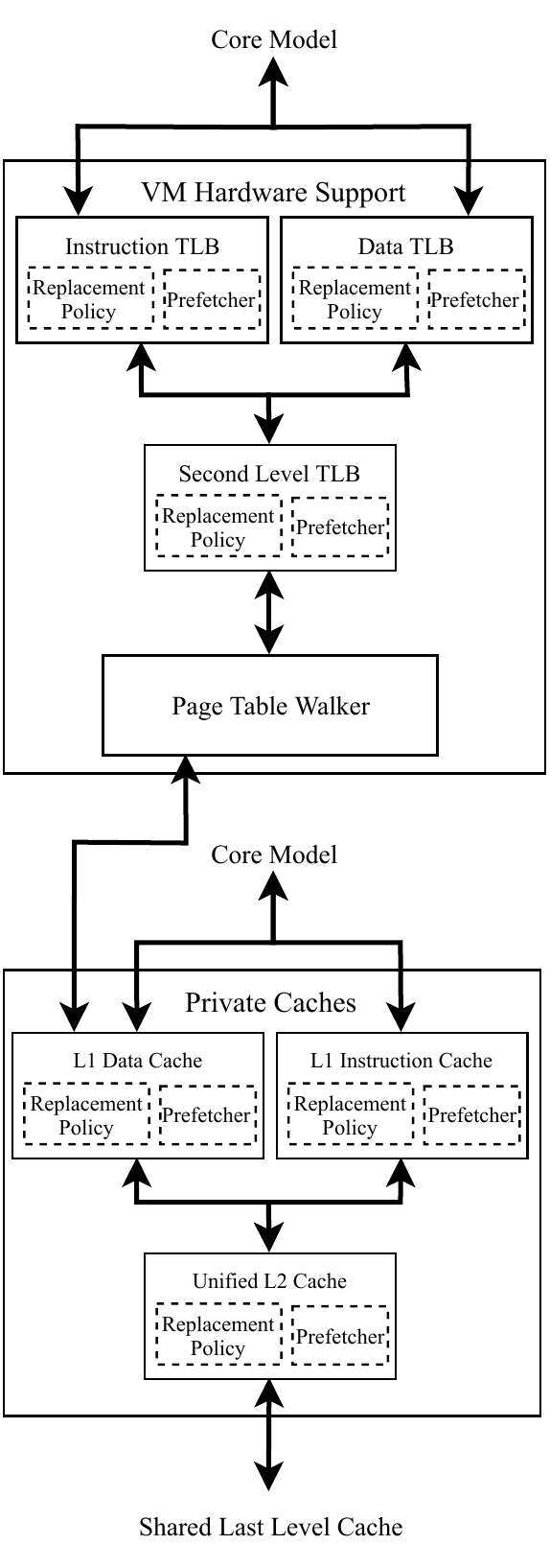}
    \hrule
    \caption{The virtual memory hardware support structures and
      private caches are instantiated for each core model. Each TLB
      and cache hierarchy level has interchangeable prefetcher and
      replacement policies.  Each core sends requests for virtual-to-
      physical address translations to its TLBs. Each core sends
      instruction or data memory requests to its private caches.  On a
      Second Level TLB miss, the Page Table Walker navigates the page
      table through a series of memory reads to find the translation.
      Page table entries are cached in the private and shared caches,
      just like any other data.}
    \label{fig:mem_model}
\end{figure}

\section{Modularity} \label{sec:modules}
Typically in computer architecture, the student or researcher
approaches their research with the goal of improving only a small part
of a large processor.  A researcher rarely attempts to develop an
entirely new processor from scratch.  Therefore, there is a need for a
highly modular simulator, whose operative elements are easily
substituted for alternative elements. A modular design permits
convenient comparisons between competing component designs, as the
underlying simulation environment is otherwise identical. This ease of
comparison is important in both an academic and an industrial
environment because it simplifies communication with interested
audiences.  It is far easier to communicate the benefit of a change
when it is readily evident to the audience that the performance gain
is attributable to the design change, and not a peculiarity of the
underlying system.

A modular design also enables easier collaboration between
researchers.  A design can be easily shared when the essence of the
design is limited to a few files of source code. ChampSim's modular
design minimizes the amount of repetitive code between modules.

ChampSim defines four types of modules: branch predictors, branch target
predictors (BTPs), memory prefetchers, and cache replacement policies.
Each cache and TLB must define a prefetcher and a replacement
policy. Prefetchers and replacement policies can be heterogeneous across
cache and TLB levels. Each core defines a branch predictor and BTP that
also can be heterogeneous between cores.  ChampSim detects all
required modules at configuration time and links a component's source
as needed.

Each module implements a set of predefined
functions. These functions are called when certain events occur in the
underlying simulator model, and are generally referred to as
``hooks.'' The hooks are implemented as member functions of either the
core or cache models to allow modules access to all structures within
their respective model. This structure permits designs that may have
unrealistic visibility into other parts of the model to allow for
extending a design's capabilities or to collect statistics.
A module's attempts to edit the internal state of the core or cache
model is considered undefined behavior and may adversely affect the
simulator's operation.  More
information is provided to the modules than some may consider
reasonably constructible in a processor. Such information is not
provided as an implication that it should be possible, but to permit
designs that make such a case.
We leave the evaluation of whether a design is
constructible to the discretion of the researcher and the committees
responsible to review the design. Figure~\ref{fig:templates} lists the
required hooks for each module in ChampSim.  All of the listed hooks
must be implemented, even if they are empty functions and perform no
action.

ChampSim is designed to be flexible and permit a wide variety of core
and cache parameters to be defined within a JSON configuration file,
with reasonable defaults as a fallback. Before compilation, a
configuration step is required in which the configuration script
examines a user-specified configuration file to determine how to build
the simulator. The core, caches, and other structures are sized
individually with values extracted from the configuration file. This
allows ChampSim to simulate not only homogeneous systems but
heterogeneous multicore systems with each core having different sized
internal and memory structures. The modules are built with compiler
flags that instruct the preprocessor to substitute the simulator hook
function names with unique, procedurally generated names.  The
configuration script generates discriminator code that ensures that
only hooks corresponding to the correct module are called at run time
by each cache and core model. ChampSim is capable of simulating
multicore systems with individual private caches and virtual memory
structures such as the one shown in Figure~\ref{fig:modularity}. The
virtual memory and cache hierarchy is illustrated in
Figure~\ref{fig:mem_model}.  This is only an example of ChampSim's
default configuration and may be reconfigured based on a user's needs.

To simulate multi-programmed workloads, at run time, the user
provides one trace file for each core, statically assigned to
that core for the duration of the simulation.  An upcoming version of
ChampSim will allow more flexibility in the trace assignments, as well
as multi-threaded trace simulation (see Section~\ref{sec:future}).

\subsection{Branch Predictors}

A hardware branch predictor is commonly included in contemporary
processors to mitigate the performance impact of branches. If a given
branch is predicted incorrectly, the subsequent wrong path
instructions must be flushed from the pipeline.
ChampSim's traces do not include any wrong path
instructions, so it instead models this with a fixed-latency penalty
following the resolution of incorrectly predicted branches. In ChampSim,
the branch predictor module hooks are member functions of the core
model. These hooks allow for designs to create prediction structures
that have access to recent branch information and branch
outcomes. The branch predictor's hooks are as follows:

\begin{itemize}
\item \texttt{initialize\_branch\_predictor()}: Each core calls this hook at the
  beginning of the simulation after the core has been fully
  initialized. This hook can be used to initialize structures in cases
  where the static initialization of C++ evaluation would occur too
  early.  Furthermore, as a member function of the core model, the
  initialization hook can observe the size of internal core
  structures.

\item \texttt{predict\_branch()}: This hook is called for each branch
  instruction in the execution path of the program.  It is responsible for
  predicting whether the branch is taken or not, and returns a boolean
  value.

\item \texttt{last\_branch\_result()}: This hook presents the branch
  predictor's opportunity to train on the true outcome of each branch
  instruction. The same information is provided to the branch 
  predictor as \texttt{predict\_branch()}, but updated with the actual
  branch outcome.
\end{itemize}

\subsection{Branch Target Predictors}

The branch target predictor (BTP) works in concert with the branch
direction predictor to attempt to fully predict the outcome and target
of branch instructions. The BTP provides the next instruction pointer
after a taken branch. Incorrect predictions of branch targets cause wrong path
execution in real processors.  As with branch direction misprediction,
ChampSim models these mispredictions with fixed-size delays. All BTP
modules are member functions of the core model. Similar to the branch
predictor module, these hooks allow for the BTP to receive
information about observed branches and their outcome. These hooks are
named based on branch target buffer (BTB) operations but can be used
to design other forms of BTPs.

\begin{itemize}
\item \texttt{initialize\_btb()}: Each core calls this hook at the
  beginning of the simulation, after the core has been fully
  initialized. This hook can be used to initialize structures within
  the BTP in cases where the static initialization of C++ evaluation
  would occur too early. Furthermore, as a member function of the core
  model, the initialization hook can observe the size of internal core
  structures.

\item \texttt{btb\_prediction()}: This hook is called for each branch
  instruction in the execution path of the program.  The BTP is
  given the instruction pointer of the branch instruction and its
  branch type. It must attempt to predict the byte address of the
  branch target and whether the branch is always taken, returning these
  values as a tuple. The BTP may indicate that the branch is predicted not taken
  by returning a target address of 0. If this disagrees with the branch
  direction prediction, the module predicting not-taken receives priority.

\item \texttt{update\_btb()}: This hook presents the BTP's
  opportunity to train on the true outcome of each branch instruction.
  This hook provides the branch instruction pointer, the true branch
  target, whether the branch was taken, and the branch type to train
  its internal structures to make better predictions.
\end{itemize}

\subsection{Data Prefetchers}

Memory prefetchers are implemented in caches to reduce miss rates by
anticipating future demand accesses. Prefetchers require information
regarding cache accesses and the accesses that result in cache
fills. This information allows the prefetcher to observe cache
behavior, generate predictions, and receive feedback regarding
predictions. In particular, memory prefetchers are assumed to operate
asynchronously with the operation of the underlying cache model,
precipitating a need for the \texttt{prefetch\_line()} callback
function.  While other modules supply their information as return
values for function hooks, a prefetcher may operate many cycles
following the initiating event.  All data prefetcher hooks are member
functions of the cache model, allowing for prefetcher developers to
extend the prefetcher's visibility of the cache's behavior as they see
fit.

\begin{itemize}
\item \texttt{prefetcher\_initialize()}: Each core calls this hook at
  the beginning of the simulation after the cache has been fully
  initialized. This hook can be used to initialize prefetcher
  structures if static initialization would occur too early.

\item \texttt{prefetcher\_cache\_operate()}: The cache calls this hook
  when a demand request occurs.  By default, this hook is only called
  for loads and prefetches from upper cache levels, though this is
  configurable. A prefetcher uses this hook to examine the access
  patterns at its cache level on a per-address basis to predict future
  cache accesses. As the primary function that communicates these
  access patterns to the prefetcher, this function also provides the
  prefetcher with the memory access's address, the instruction address
  that caused the memory access, whether the access resulted in a hit
  in the cache, and the type of memory access. The prefetcher also
  receives metadata from the incoming access that may be provided by a
  prefetcher in an upper cache level. Likewise, the prefetcher may
  return metadata to the cache to embed within the memory access's
  packet to communicate with lower levels of the cache hierarchy.

\item \texttt{prefetcher\_cache\_fill()}: This hook is called each
  time a block is filled in the cache.  Its parameters include
  information about which blocks were evicted and filled.  Prefetchers
  can use this hook to evaluate their own accuracy, estimate cache
  miss latency, or to make future prefetching decisions about evicted
  blocks.

\item \texttt{prefetcher\_cycle\_operate()}: This hook is called once
  every cycle on the same clock frequency as the underlying cache.
  This hook enables developers to precisely simulate the pipelining of
  complex prefetchers, if desired.  Memory prefetch timing is often
  crucial to avoid misses, so complex prefetchers may use this
  function to emulate the delay between a prefetch and its initiating
  event.

\item \texttt{prefetcher\_final\_stats()}: Once all instructions are
  executed, this hook prints user-defined statistics recorded
  throughout the simulation to the simulation's output.

\item \texttt{prefetch\_line()}: This hook allows for the asynchronous
  operation of the prefetcher.  Unlike other hook functions, this
  function is never called by the underlying simulator but must be
  called by the user to perform a prefetch. The prefetcher might call
  this in response to a demand request, a fill, or on a particular
  cycle, and can be called as many times as the user wants. The user
  can choose whether to fill into either the prefetcher's current
  cache level or a lower cache level.
\end{itemize}

\subsection{Instruction Prefetchers}

Unlike the data prefetcher modules, the instruction prefetcher module
hooks are members of the core model and are connected to the L1
instruction cache.  Most instruction prefetcher hooks are called by
the cache when certain events occur in the cache.  Instruction
prefetchers have the same hooks as the data prefetchers, and are
called on the same events. The only exception is the
\texttt{prefetch\_line()} function, which is replaced by
\texttt{prefetch\_code\_line()}. Furthermore, instruction prefetchers
have one additional hook:

\begin{itemize}
\item \texttt{prefetcher\_branch\_operate()}: This hook is called when
  a branch instruction is read from the trace.  This hook can give the
  instruction prefetcher information about the boundaries of basic
  blocks.
\end{itemize}

\subsection{Replacement Policies}

Cache replacement policies choose which cache line to evict from a
cache set.  Many do this by using a heuristic to predict the future
utility of cache lines, and evict the line with the lowest expected
utility.
The information provided to the replacement policy allows it to view
the cache's demand access stream.
All of the hooks of a replacement policy are member
functions of the cache model.

\begin{itemize}

\item \texttt{initialize\_replacement()}: Each core calls this hook at
  the beginning of the simulation, after the cache has been fully
  initialized. This hook can be used to initialize structures specific
  to the replacement policy where static initialization would occur too
  early.

\item \texttt{find\_victim()}: The cache calls this hook when a cache
  set requires a victim cache block.  The module returns the cache way
  within the set to be replaced. Cache bypassing is also an option,
  and is indicated by returning a value equal to the maximum number of
  ways.

\item \texttt{update\_replacement\_state()}: The cache calls this hook
  to update the replacement policy on a change in the cache's state
  due to a cache hit or fill.  The information provided to the
  replacement policy enables the replacement structures to receive
  feedback on whether their victim selection was harmful to the
  system's performance.

\item \texttt{replacement\_final\_stats()}: This hook is called at the
  end of the simulation after all instructions are executed.  This
  hook is used to print user-defined statistics recorded during the
  simulation.
\end{itemize}

%% file: application.tex
\section{Application} \label{sec:application}

ChampSim has already found a place in the ecosystem of computer
architecture simulation as a lightweight tool to introduce ideas to
the community.  It has promoted new and innovative research, fueled in
part by industry-sponsored competitions.  Furthermore, it shows
promise as a tool for classroom use, striking a balance between being
simple and accessible, and being a feature-rich simulator.

\subsection{Research}

The rapid prototyping and evaluation of computer hardware designs is
essential to the continuing development of computer architecture.
Established industry heavyweights design and implement simulators for
internal use.  These proprietary simulators are rarely disclosed,
since doing so may reveal valuable trade secrets that will hinder them
competitively.  Therefore, the research community must itself develop
and support open-source simulators to conduct independent computer
architecture research.


ChampSim has already been in use among the community for some time.
Due to its exposure through computer architecture competitions,
several high-impact publications have founded their results on ChampSim.
It has been key in demonstrating the excellence of many memory
prefetchers, both of
data~\cite{kalani2021cal,zhang2020micro,kim2017asplos,kim2016micro,bakhshalipour2019hpca,bhatia2019isca,wu2021itc,wu2019micro,wu2019isca,pakalapati2020isca}
and instructions~\cite{ros2021isca}.  It has found use in
evaluating cache replacement policies, particularly in the last
level cache~\cite{sethumurugan2021hpca,peneau2018springer,deng2021date,jain2018isca}.
ChampSim's view of a TLB as functionally identical to a cache has led to
research into prefetching in
TLBs~\cite{vavouliotis2021micro,vavouliotis2021isca}. The modularity
of ChampSim's branch predictor has shown benefits in novel
works~\cite{lin2019arxiv}. The branch target buffer has
also been studied with ChampSim~\cite{asheim2021cal}.

As a fast, lightweight simulator, ChampSim's relatively small code
footprint has proven to be easy to understand and modify.  While some
have leveraged the module system to develop novel works, others have
additionally modified the underlying model
itself~\cite{ishii2021ispass,kumar2021cal,barboza2021hpca,shi2019micro}. Users
can furthermore modify the existing program tracer to generate traces
of different formats to meet particular needs.

In only a few years, ChampSim has shown to be a valuable tool for
academic research.  As of this writing, research performed using
ChampSim has already received 352 citations. Development of ChampSim
is still ongoing, and new features are continually being added.
ChampSim is likely to foster further work in these areas as the
simulator's development continues.

\subsection{Competition}

There has been a rise in industry-sponsored contests to promote memory
system research.  ChampSim has been featured in:

\begin{itemize}

\item The Second Data Prefetching Championship (2015)\cite{dpc2} 
\item The Second Cache Replacement Championship (2017)\cite{crc2} 
\item The Third Data Prefetching Championship (2019)\cite{dpc3} 
\item The First Instruction Prefetching Championship (2020)\cite{ipc1} 
\item The ML-Based Data Prefetching Competition (2021)\cite{mldpc} 

\end{itemize}

\begin{table}
  \centering
  \begin{tabular}{cc}
    Competition & Submissions\\\hline
    DPC-2 \cite{dpc2} & 13\\
    CRC-2 \cite{crc2} & 15\\
    DPC-3 \cite{dpc3} & 14\\
    IPC-1 \cite{ipc1} & 16\\
    MLDPC \cite{mldpc} & 4\\\hline
    Total & 62
  \end{tabular}
  \caption{The number of submissions to a selection of competitions. Each of these submissions is a unique proposal in the space of computer architecture.}
  \label{tab:submissions}
\end{table}

In these competitions, a call for submissions solicits submissions
of innovative designs for a particular module.  The
submissions are compared and ranked by metrics such as instructions
per cycle or cache misses.  These contests encourage many designs to
be compared and rewarded for their merits.  They have increased the
rate of publication in the field of memory prefetching and cache
replacement.  There have been many submissions to these competitions,
listed in Table~\ref{tab:submissions}, which have produced many
follow-up publications.  ChampSim has been the simulator of choice in
these competitions due to its availability, being open-source, ease of
use, requiring minimal dependencies, and speed, evaluating billions of
instructions per hour.

ChampSim was designed from the beginning to make these kinds of
comparative contests simple to organize, evaluate, and participate in.
Contestants typically are given 2 months to ramp up with ChampSim and
produce novel work. The high rate of participation in each contest is
evidence that ChampSim is accessible and easy to use. The modular design
creates a competition workflow wherein contestants need only submit
the source code for the module under evaluation. The competition
organizer can then easily compile, run, and evaluate the competitors'
submissions.  We invite any others who would like to organize
competitions in any field of CPU architecture to contact us.

\subsection{Education}

In a classroom setting, instructors are given approximately fourteen
weeks to teach students about tools needed to complete assignments
while also teaching them about the topics at hand.  It is accordingly
difficult to use a heavyweight simulator
in a classroom setting.  Students may be
asked asked to configure and install systems they do not
understand for the purpose of only a few assignments.  If this
configuration and installation is difficult, it impedes the students'
progress by increasing their frustration and decreasing their
engagement with the classroom topics.  As before, the tools used in an
academic setting should, as much as possible, lower the barrier to
entry for these young entrants to the field.  In such a scenario, a
lightweight, easy to understand simulator is preferred.  ChampSim
meets the needs of educators and 
allows students to implement microarchitecture techniques they learn
in class.

ChampSim has been designed such that the only dependencies are
up-to-date C++ and Python environments.  Further dependencies
could increase the likelihood that a student may encounter
difficulties when initializing their simulation environment.
Its development target includes compilers available
natively in popular long-term supported operating systems that are
commonly available to students.  At the time of writing, this includes
GCC 7, Clang 4, Microsoft Visual Studio 2017, and Python 3.6.

ChampSim has a small codebase, approximately 5000 lines of code, and so
it is possible to comprehend in its entirety.  The design of ChampSim
does not necessarily attempt to precisely replicate all hardware
functions, but to emulate it in a readable way.  These design
decisions allow a student to rapidly grasp its functionality,
configure it, and implement or modify techniques they've learned about
in class, all within the span of one academic semester.  ChampSim has been
successfully used in classroom settings at Texas A\&M University at both the undergraduate and graduate levels.

In the undergraduate level course, one the students' first assignments
was to download ChampSim and to become familiar with each of its
components.  The students' final project was to optimize a last-level
cache replacement policy, and prove that their technique outperformed
the base case (base cases were different for each student).  All
students were able to successfully complete the first assignment and
never during the semester-long project faced difficulties finding the
computer resources to download and configure ChampSim.  As the semester
progressed, students were able to identify which module of the
simulator modeled each of the components discussed in the course and
were able to visualize the topics by studying and modifying the
simulator.  Students maintained a high level of interest during the
lectures, were able to run simulations and correctly interpret the
results, and showed a high level of mastery on the course exams.

ChampSim was also used in a graduate course at Texas A\&M University.
In particular, the focus of this course is in out-of-order core and
memory system microarchitecture.  As this is an introductory course in
microarchitecture research, the aim is to get the incoming students up
and running on a simulator as soon as possible, so that they can use
most of the semester developing a course project.  To this end,
ChampSim was used in two early semester assignments, one exploring the
impact of different branch predictors on performance of cores with
different OoO width, and another exploring the impact of differing
prefetch and replacement policies on the L2 cache.  After this
introduction, nearly all the course students choose to use ChampSim for
their course projects, and those that did were more likely to achieve
their project goals by the end of the semester.

For an undergraduate computer architecture course at Texas A\&M International University, the students used ChampSim throughout the course for project work.
At the beginning of the semester, students were asked to download and set up the simulator.
All students were able to do in a matter of minutes, despite none of these students having previous experience with any other simulator.
For their first assignment, students were asked to run simulations changing the configuration of the simulator testing different cache sizes, different replacement policies, and other configurable aspects, and were asked to reason about results obtained.
As the semester progressed and topics were covered during lecture, being able to plainly read the source code of ChampSim helped the students understand the concepts in the course and how the concepts could be implemented in real systems.

In the lecture on memory hierarchy, students were asked to implement their own replacement policy and, while it was not expected nor required for such policy to show a performance improvement, it was required for the student to justify their idea and implementation given what they learned in class.
The same kind of assignment was given over branch prediction and memory prefetching.
These types of assignment challenged the students to go beyond the concepts presented to them on their textbook and to think outside the box about the details in the implementation of such components.
ChampSim was shown to be a powerful supplementary teaching tool for this type of course.

We continue to advocate for its usefulness as a tool to invite new students into the domain of computer architecture research.
The authors are aware that ChampSim has also been used at additional universities at both the undergraduate and the graduate level, each with reports of success and discovery on the part of the students.
We hope that ChampSim continues to serve as a low-impendance tool that permits discovery and research with a low degree of frustration.

%% file: related_works.tex
\section{Related Works}
\label{sec:related_works}
This section provides a brief overview of selected prior works in
architectural simulation and evaluation with regards to the key design
principles discussed in Section~\ref{sec:design}. Simulation and
emulation is the main tool used by computer architects in academia to
evaluate the performance impact of new microarchitectural designs. As
a result, simulation methodology and simulator design is exceedingly
broad area of study and development~\cite{akram2019access,
  brais2020survey}.  As such, there have been an extremely large
number of simulators proposed over the years, here we discuss a subset
of the most popular current simulators.

Scarab~\cite{scarabsimulator} is a recently developed simulator
capable of performing cycle-accurate execution-driven simulation by
using Intel's PIN to execute application binaries passed by a user.
For faster simulation times, Scarab also supports trace-driven
simulation. The simulator models an out-of-order pipelined processor,
a full cache hierarchy, and interfaces with
Ramulator~\cite{kim2016ramulator} to provide an accurate and highly
configurable main memory model. Scarab is a promising simulation
framework with a low start-up time, but users are required to use
Intel-based systems to use the Pintool based execution-driven
mode. Scarab is suitable for more advanced architects' use and may not
be appropriate for a classroom setting with students of varying
expertise.

Gem5 is a popular event-driven simulator with a wide variety of
features to enable the exploration of most areas of microarchitectural
design. A researcher that finds gem5 does not meet their needs will
likely find another work that extends gem5 for their
purpose~\cite{gubran2019isca,shao2016micro,power2015cal,qureshi2021taco}.
Gem5's full system mode allows users to emulate a complete computing
system with devices and for an OS to be executed on the emulation. For
designs that do not require OS-support, gem5 provides system emulation
mode in which a user may provide a binary for the simulator to
execute. The simulator supports most available ISAs and is extremely
modular, permitting researchers to plug-and-play system components,
interchange cache hierarchy organizations, experiment with various
network on chip topologies, and explore various core models. This high
amount of configurability allows gem5 to meet most architectural
research needs.  Still, this robustness and configurability leads to
very high complexity which heavily increases the barrier of entry and
startup time for new students and researchers.

Zsim~\cite{sanchez2013zsim} is a multicore simulator that breaks the
simulation down into multiple phases, relying on Pin to perform
instrumentation of an input binary to relieve the burden of highly
accurate timing models. Each core modeled in the system is executed in
a separate thread based on the instrumentation phase to allow for
highly scalable system simulation. The simulator is configurable and
extendable to allow researchers to represent a wide variety of memory
hierarchies and heterogeneous systems. Similar to Scarab, zsim limits
the host system to use an Intel processor subsequently limiting the
system to the x86 ISA. Due to the simulation methodology used, the
host system may require specific configurations that a student may not
have permission to modify.

Sniper~\cite{carlson2011sniper} is a highly parallel multicore 
simulator that extends Graphite~\cite{millerhpca2010} to add interval model 
implementation. This methodology allows for faster simulation by abstracting 
the system model to events that affect timing instead of tracking individual 
instructions as they traverse the pipeline. The simulator supports 
heterogeneous core models and operating system execution. The developers 
provide Python-based scripts to control and monitor the simulation during 
runtime. Similar to Scarab and zsim, PIN dynamically executes user-inputted 
binaries. Users are required to request access to the latest version of 
Sniper's source code. Overall, the simulator is configurable and supports 
multiple system models, but is limited to x86 execution. The complexity of 
interval modeling makes the system difficult for novice architects and students 
to modify, resulting in a high barrier of entry. 

Each of the simulators discussed here has been widely used for a
variety of research applications and resulted in a large number of
publications. These tools provide configurable features but require
extensive knowledge and ramp-up time to begin design exploration.
ChampSim is not an attempt to replace these tools, but rather to fill
the gap in computer architecture tools for architects. ChampSim is ideal
for users that do not require heavyweight simulation for their
research and educators seeking to provide meaningful curriculum for
semester-long courses.

%% file: future_work.tex
\section{Future Work} \label{sec:future}

ChampSim is a comparatively new tool and remains in very active
development, with new features and bug fixes being regularly added,
within the constraints of the design principles.  Because of the
growing adoption of ChampSim as a tool for education and research,
backwards compatibility is a high priority in any changes that are
made.  For example, module hook signatures are unlikely to change
(without sufficient cause) to ensure that already published artifacts
remain compatible with newer versions of ChampSim.

Nevertheless, the trace format used imposes some restrictions on the
correctness of ChampSim's simulation.  The current trace format uses
fixed-size arrays to represent register and memory location usage
information.  This may be inefficient, leaving some gaps of wasted
memory if one instruction accesses fewer memory locations or registers
than other instructions, but it also carries implicit assumptions
about which instruction set architecture is being traced.  A planned
future instruction trace format will use variable-sized arrays to
represent this information and should also use compression as part of
the trace specification.

As part of this new trace format effort, more information will be
included.  ChampSim currently views instructions as being in one of the
following classes: branch, load, store or arithmetic.  It handles all
arithmetic instructions identically, with the same arbitrary delay
imposed on each.  In the future, more accurate simulation is possible
with a more precise set of delays.

The current implementation of ChampSim assumes a workload of multiple,
single-threaded programs.  That is, each core is assumed not to
interact with any other core's execution, except to create pressure on
any shared caches.  Lock contention is a frequent source of execution
delays in multi-threaded workloads, and adding support for such
contention will improve simulation accuracy.  In the future, ChampSim
will support multi-threaded workload simulation, adding support for
synchronization constructs such as locks and barriers.  In a new trace
format, the interaction between threads can be encoded.

ChampSim currently makes no effort to maintain cache coherence and only
models non-inclusive cache hierarchies. A future step will be to 
implement other forms of clusivity and allow for cache coherence 
modeling, since an ongoing body of work, is the interaction of memory 
prefetching with various forms of each. The modular design of ChampSim
positions it well to examine the interactions of combinations of such
works, since the individual components can be trivially substituted
for one another.

Development has continued on the front end of ChampSim since the First
Instruction Prefetching Competition \cite{ipc1} to make the front
end increasingly fetch directed.  The branch direction and target
predictors may, as a part of this effort, become asynchronous like the
memory prefetchers, allowing for prediction latencies longer than a
single cycle.

%% file: conclusion.tex
\section{Conclusion}

In this work, we have discussed ChampSim, an architectural simulation
tool designed for ease of use and rapid simulation.  We have presented
it as a tool for research, competition, and education, specifically
that these may be more accessible to an upcoming cohort of new
architects.  In this way, we hope to promote a broad, diverse, and
inclusive community of computer architecture.

The design of the simulator focuses on a low startup time, such that
the user is able to begin using ChampSim with minimal frustration; broad
applicability, so that any new researcher is able to comprehend their
place in the broader scope of the environment; and configurable
design, where the user is able to simulate their work in a variety of
contexts with ease.  ChampSim accomplishes this with a modular design.
Each element of the simulator is designed to be replaceable with an
equivalent element.  This allows for prompt and proper comparison
between competing designs.

ChampSim has been used in a variety of research and teaching endeavors
throughout its lifetime.  It has been shown to have a strong impact on
the community by promoting novel works in the field.  The competitions
that have been centered around ChampSim have prompted many submissions and led to many high-impact publications.
ChampSim has also shown to be useful in a classroom, as a tool where
students can explore and develop their ideas.  We believe that this
work has already found a niche in the landscape of architectural
simulation, and we hope that this work can continue to grow and to
serve the community well.